\documentclass[prl,showpacs,twocolumn,
floatfix]{revtex4}
\usepackage{graphicx,amsmath,amssymb,latexsym,amsfonts,subfigure,bbm}
\begin{document}
\title{Shot noise in the chaotic-to-regular crossover regime}
\author{Florian Aigner}
\author{Stefan Rotter}
\author{Joachim Burgd\"orfer}
\affiliation{Institute for Theoretical Physics, Vienna University of
Technology, A-1040 Vienna, Austria}
\date{\today}
\begin{abstract}
We investigate the shot noise for phase-coherent quantum transport in the
chaotic-to-regular crossover regime. Employing the
Modular Recursive Green's
Function Method for both ballistic and disordered two-dimensional cavities 
we find the Fano factor and the
transmission eigenvalue distribution for regular systems to be surprisingly
similar to those for chaotic systems. We argue that 
in the case of regular
  dynamics in the cavity, diffraction at the lead openings is the dominant
  source of shot noise. We also explore the onset of the crossover from
  quantum to classical transport and develop a quasi-classical transport model
  for shot noise suppression which agrees with the numerical quantum data.
\end{abstract}
\maketitle
The significance of noise induced by the discreteness of the electron charge
(``shot noise'') first investigated almost a century  ago \cite{schottky} has
resurfaced in the field of {\it mesoscopic physics}
\cite{been1,blanter1}. Shot noise carries information about the crossover from a deterministic (classical) particle picture of 
electron motion to a probabilistic (quantum) description, where electrons
behave as matter waves. The uncertainty inherent in a quantum
picture gives rise to noisy transport. In conductance through quantum dots the
correlations between electrons in the Fermi sea 
lead to a suppression of shot noise $S$ relative to the Poissonian value
of uncorrelated electrons $S_P$ \cite{been2}. The
reduction relative to the completely random limit is customarily expressed in
terms of the Fano factor $F=S/S_P$.\\Most 
investigations to date have focused on quantum dots whose
classical dynamics is fully chaotic \cite{been2,tworz,agam,jalabert,sylvestrov,jacquod,naz,sukhorukov}. In this limit, random matrix
theory (RMT) \cite{jalabert} predicts a universal value for the Fano factor,
$F=1/4$. The applicability of this RMT result requires, in addition to the
underlying chaotic dynamics, dwell times in the
open cavity $\tau_D$ which are sufficiently long compared to the Ehrenfest
time  $\tau_E$. The latter estimates the
time for the initially localized quantum wavepackets to spread all over the
width $d$ of the cavity (typically $d\approx\sqrt{A}$ with $A$ area of the 
dot) due to the divergence of classical chaotic trajectories. 
It can be estimated as \cite{zaslavsky}
\begin{equation}\label{eq:ehrtime}
\tau_E = {\Lambda^{-1}} \ln (d/{\lambda_F})\,,
\end{equation}
where $\Lambda$ is the Lyapunov exponent $(\Lambda > 0 $ for a chaotic
cavity), and $\lambda_F$ is the de Broglie wavelength associated with the
wavenumber at the Fermi surface $k_F$. The limit 
$\tau_E/\tau_D\ll 1$ corresponds
to the quantum (or RMT) regime and $\tau_E/\tau_D \gg 1$ corresponds to the
classical limit for which $F=0$ is expected. For
ballistic cavities in the crossover between these two regimes a simple
conjecture for $F$  was put forward
\cite{agam},
\begin{equation}\label{equ:expdecay}
F\!=\!{1}/{4}\,\exp\left(-{\tau_E}/{\tau_D}\right)\,.
\end{equation}
For cavities with a short-ranged disorder potential, an alternative crossover
behavior,
\begin{equation}\label{equ:algebraicdecay}
F\!=\!{1}/{4}\,\left(1+\tau_Q/\tau_D\right)^{-1}\,,
\end{equation}
was proposed \cite{oberholzer1,oberholzer3}, where $\tau_Q$ is a
characteristic scattering time within which the wavepacket is scattered into
random direction and thus explores the entire cavity.
$\tau_Q$ and $\tau_E$ are closely related to
another as both denote the characteristic time scale for spreading of the
wavepacket by chaotic scattering either at the boundary ($\tau_E$) or the
interior ($\tau_Q$) of the cavity. Moreover, for short ranged disorder with a
correlation length $l_C<\lambda_F$, $\tau_Q$ incorporates, just as
$\tau_E$, quantum effects and depends on an effective $\hbar_{\rm eff}$ of the
system. The
crossover from the chaotic to the regular regime is therefore
predicted to be controlled by a single ratio $\tau_E/\tau_D$ or
$\tau_Q/\tau_D$ which will be a function of both the size of quantum effects
($\hbar_{\rm eff}$) and the mean rate of irregular (chaotic) scattering,
$\langle\Lambda\rangle$. The chaotic-to-regular crossover corresponds to the
limit $\langle\Lambda\rangle\rightarrow 0$, while the quantum-to-classical
limit involves $\hbar_{\rm eff}\rightarrow 0$.\\An 
obvious test of different
predictions would be to simulate phase-coherent scattering processes on a
computer. However, conventional numerical techniques suffer from a slow
convergence rate for large $\tau_E$ and, in particular, for small $\hbar_{\rm
  eff}$. To circumvent this
difficulty an ``open'' dynamical kicked rotator model was recently
successfully used to mimic chaotic scattering in a 1D-system
\cite{tworz,jacquod}. 
Experimental tests have, so far, been limited to the regime
of $\tau_Q/\tau_D<1$ \cite{oberholzer1}.\\One 
open question not yet well understood is the behavior of shot noise
for motion in a regular rather than chaotic cavity, i.e.~in the
limit $\Lambda \rightarrow 0$. For a mixed system lower values of $F$ have
been observed \cite{sim} suggesting that for regular systems $F$ may vanish. 
Taken at face value, Eq.~(\ref{equ:expdecay}) yields $F=0$
for the case of $\tau_E \rightarrow
\infty$ or $\Lambda \rightarrow 0$ at fixed value of $\hbar_{\rm eff}$. This
result would correspond to a complete suppression
of shot noise.
On the other hand, the notion of noisy transport as a result of
the wave nature of electrons would suggest that noise should be 
inherent in phase-coherent
transport through mesoscopic structures irrespective of the chaoticity of the
underlying classical dynamics. In the present letter we analyze 
a model for transport through a cavity
that allows to investigate the crossover regime from chaotic to regular
dynamics, i.e.~$\Lambda \rightarrow 0$, and the onset of the crossover 
from quantum to classical dynamics.
In our model, the dwell time and the mean rate of chaotic spreading of the
wavepacket, $\langle\Lambda\rangle$, can be tuned independently.
Surprisingly, we find that the Fano factor and the distribution of
transmission eigenvalues for regular systems resemble closely that of 
chaotic systems. We interpret this observation in terms of the ubiquity 
and dominance of diffractive scattering in phase-coherent 
cavities.\\\begin{figure}[!t]
\begin{center}
\includegraphics[angle = 0, width = 85mm]{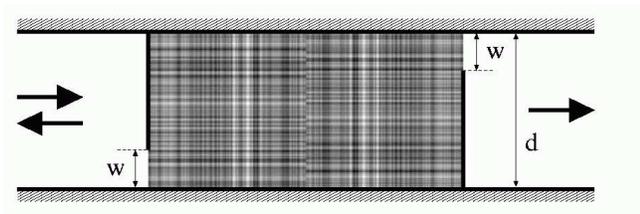}
\caption{Rectangular quantum billiard with tunable shutters and
tunable disorder potential (gray shaded area).  
Electrons are injected from the left into the cavity 
region of size $A=2d^2$, width $2d$ and height $d$.
Tuning the opening of the shutters $w$ and the strength of the disorder
potential $V_0$, the onset of the crossover from quantum-to-classical
and chaotic-to-regular scattering can be investigated,
respectively.}
\label{fig:1}
\end{center}
\end{figure}We choose
first a scattering geometry (Fig.~\ref{fig:1}) 
which consists of a rectangular cavity to which two 
leads of width $d$ are attached via tunable shutters with an opening
width $w$. To reduce  
direct transport between the shutter openings, they are placed
on the bottom and top end of the leads, 
respectively. The cavity region of width $d$ and length $2d$  
contains a disorder potential $V$ characterized by its mean value 
$\langle V \rangle =0$, and the
correlation function 
$ \langle V(x) V(x+a) \rangle = \langle V^2 \rangle \exp(-a/l_C)$ 
\cite{koschny}. 
The correlation
length $l_C$ is typically a small fraction of the Fermi wavelength
$l_C/\lambda_F \approx 0.12$ and the potential strength 
$V_0=\sqrt{\langle V^2\rangle} $ is weak,
$V_0/E_F\le 0.1\,$. The dependence of
$\tau_D$ on the shutter opening was
determined by a Monte-Carlo sampling of $\sim 10^5$ classical trajectories,
$\tau_D \approx 2.66/(wk_F)$, i.e.~approximately independent of
$V_0$. In the limit of vanishing disorder
potential $(V_0 \rightarrow 0)$ the motion inside the cavity 
becomes completely regular.\\Our quantum calculation proceeds within 
the framework of the modular
recursive Green's function method (MRGM) \cite{rotter1} 
which allows to treat
two-dimensional quantum dots with relatively small $\lambda_F$ (or small
$\hbar_{\rm eff}$). 
The MRGM requires, however, separable modules. We construct the
latter by decomposing the cavity region into two square modules for each of
which we choose a separable random potential $V(x, y)=V_1 (x)+V_2(y)$. In
order to destroy the unwanted separability and to ensure chaotic dynamics, 
we build up the cavity by combining two
identical modules, however, rotated by $180^{\circ}$ relative to each
other (see Fig.~\ref{fig:1} for illustration). 
As the Fermi wavenumber and the 
width of the leads $d$ is independent of the shutter openings 
$w$, this device allows to 
study two different crossovers of shot noise:
(1) By changing the opening of the shutters $w$ the dwell time
in the cavity $\tau_D$ can be tuned without changing the
Fermi energy $E_F$. 
With the increase of $w$ the onset of the quantum-to-classical crossover can
be probed, with the sideffect, however, that 
not only $\hbar_{\rm eff}$ is reduced but simultaneously the classical
phase space structure is altered. This closely resembles the
parameter tuning in the experiment \cite{oberholzer1}. (2) By
tuning the strength of the disorder potential, the rate of chaotic spreading,
$\langle\Lambda\rangle$, and thus $\tau_E$ (or $\tau_Q$, used in the following
interchangeably) is varied.\\We evaluate
the transmission amplitudes $t_{mn}$ for an electron injected from 
the left by projecting the Green's function at the Fermi 
energy $G(E_F)$ onto all modes $m,n\in[1,\ldots,N]$ in the in- and 
outgoing lead, respectively. The Fano factor $F$ is then calculated from the
$N$-dimensional transmission matrices $t$ \cite{blanter1},
\begin{equation}
F = \frac{\langle\rm{Tr}\; t^\dagger t (\mathbbm{1}-t^\dagger t)\rangle}
{\langle\rm{Tr} \; t^\dagger t\rangle} =
\frac{\langle\sum_{n=1}^{N} T_n (1-T_n)\rangle}
{\langle\sum_{n=1}^{N} T_n\rangle}\,,
\end{equation}
with $T_n$ being the eigenvalues of $t^\dagger t$. The brackets 
$\langle\ldots\rangle$ indicate that we average over 150 equidistant
points in the wavenumber-range $k_F\in[40.1,40.85]\times\pi/d$,
where 40 transverse lead modes are open. Note that this mode number is higher
than in previous studies. We choose a nearest-neighbor 
spacing $\Delta x=\Delta y$ in the Cartesian discretization grid 
such that the Fermi wavelength is well resolved by a large number of
grid points, $\lambda_F\approx 32 \Delta x$. With these settings we have 
a total number of
$\sim 8.5\times 10^5$ grid points in the interior of the 
cavity.\\\begin{figure}[!t]
\begin{center}
\includegraphics[angle = 0, width = 85mm]{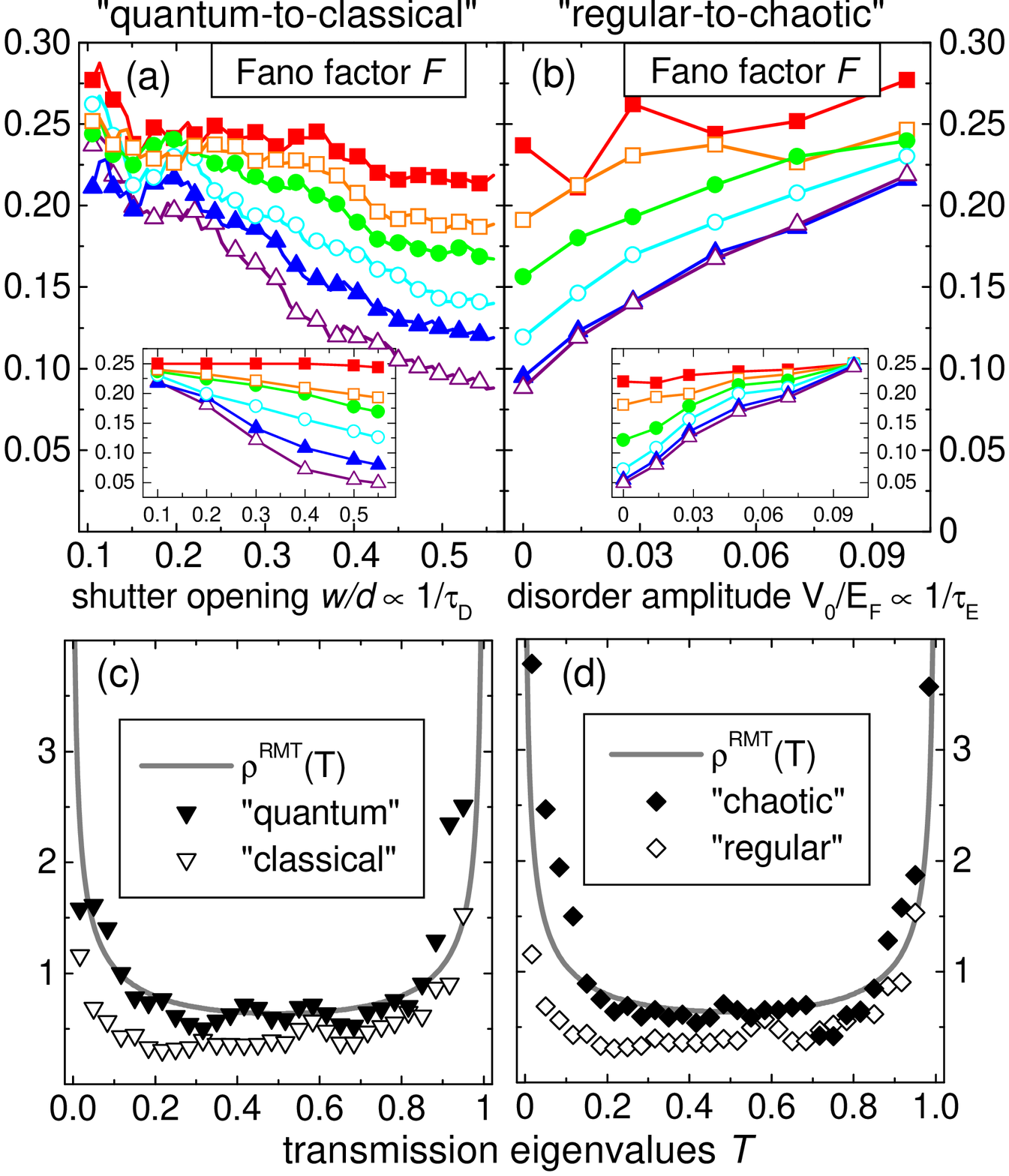}
\caption{(a) Fano factor $F$ in the quantum-to-classical
crossover regime for the cavity depicted in Fig.~\ref{fig:1}. 
Curves shown correspond to different strengths of the
disorder potential: $V_0/E_F = 0.1\,({\scriptstyle \blacksquare}), \;0.07\, 
({\scriptstyle\square}),
\;0.05\, (\bullet), \;0.03\, (\circ), \;0.015\, (\blacktriangle), \;0\,
(\vartriangle\nobreak)$. (b) Regular-to-chaotic
crossover regime of $F$. Curves corresponding to different shutter opening
ratios are shown: $w/d = 0.55\,({\scriptstyle
\blacksquare}),
\;0.5\, ({\scriptstyle \square}),\;0.4\, (\bullet), \;0.3\, 
(\circ), \;0.2\,
(\blacktriangle), \;0.1\,(\vartriangle\nobreak)$. Insets of (a) and (b)
depict the fit parameter free prediction with diffractive corrections
[Eq.~(\ref{eq:integralfano})]. (c) Distribution of transmission 
eigenvalues $\rho(T)$ in the quantum-to-classical crossover: 
$V_0/E_F=0\,(\blacktriangledown),0\,(\triangledown)$ and
$w/d=15\,(\blacktriangledown),50\,(\triangledown)$. (d)
$\rho(T)$ in the chaotic-to-regular crossover: 
$V_0/E_F=0.1\,(\blacklozenge),0\,(\lozenge)$ and
$w/d=50\,(\blacklozenge),50\,(\lozenge)$.}
\label{fig:2}
\end{center}
\end{figure}Figure
\ref{fig:2} displays the Fano factor as a function of the inverse dwell time
$\tau^{-1}_D$ (Fig.~\ref{fig:2}a)
and the inverse Ehrenfest time  $\tau^{-1}_E$ (Fig.~\ref{fig:2}b). 
For $\tau^{-1}_D
\rightarrow 0$ (i.e.~large dwell times) $F$ approaches the universal value 1/4
irrespective of the strength of the disorder potential $V_0$, 
while for shorter dwell
times $F$ falls off gradually (Fig.~\ref{fig:2}a). 
The steepness of this
decrease is clearly dependent on $V_0$ and thus on the mean scattering rate
$\langle\Lambda\rangle$. Most striking
is the feature that for $V_0 \rightarrow 0$ 
but long dwell times the shot noise reaches 
the RMT value even though the
dynamics is now entirely regular (see Fig.~\ref{fig:2}a). In Fig.~\ref{fig:2}b
this feature is reflected by the fact that the Fano factor $F$
does not decay to zero even as $V_0 \rightarrow 0$ (i.e.~$\tau_E^{-1}
\rightarrow 0)$.
There is no compelling a priori reason why
the RMT prediction should be applicable to regular systems. 
This observation suggests that the conjectures 
[Eq.\ (\ref{equ:expdecay}) or Eq.\ (\ref{equ:algebraicdecay})]
require a modification to properly account for the shot noise in the
regular limit. We argue that the
key point is the wavepacket diffraction
at the cavity openings which has to be incorporated in the 
theoretical description of shot noise. Note that this feature  
is inherent in quantum transport and independent of the underlying
regular or chaotic dynamics \cite{rotter1}. 
Scattering due to {\it chaotic} dynamics, which lies at the 
core of RMT, certainly leads to
wavepacket spreading but does not constitute the only or, in general,
dominant source.\\Diffraction at the cavity openings
has been studied in detail \cite{wirtz1,wirtz2} and 
can be described by a standard Fraunhofer diffraction analysis for
electrons which enter or leave the cavity \cite{wirtz1}. 
An estimate for wavepacket
spreading can be found by considering the characteristic angular
injection patterns of the transverse 
modes in these openings. Averaging the individual patterns over the
total number of modes in the shutter openings,
$M={\rm int}(k_Fw/\pi)$, we obtain for 
the variance of the injection angle $\theta$
the dependence: 
$\langle\sqrt{\langle\theta^2\rangle-\langle\theta\rangle^2}\rangle
\approx 0.5\times M^{-0.5}$.
This result now enters our considerations on a modified estimate
for $\tau_E$ (or $\tau_Q$). We perform a quasi-classical Monte-Carlo transport
simulation \cite{burgl} in which we follow an ensemble of classical
trajectories subject to a random Poissonian scattering process. For the latter
we calculate the transport mean free path ($\tau_S\cdot v_F$) and the
differential scattering probability ($P(\theta)\sim d\sigma/d\theta$) in first
Born approximation, thus taking into account quantum diffractive scattering
(for $l_C\cdot k_F<1$) along the lines of Refs.~\cite{oberholzer3,
  oberholzer1}.
In this simulation, spreading results from both the injection of classical
trajectories with an initial angular distribution given by Fraunhofer
diffraction at the shutter opening and multiple scattering inside the
cavity. This allows to identify a modified
Ehrenfest time $\widetilde{\tau}_E$ as the time it takes for the ensemble to
acquire a mean spread of the order of the 
cavity width $d$ under the influence of both 
disorder (or chaotic) scattering inside the cavity as well as diffraction at
the shutter openings. At $\widetilde{\tau}_E$ the mean separation reaches 
$\Delta
r(\widetilde{\tau}_E)=\langle(\vec{r}-
\langle\vec{r}\,\rangle)^2\rangle^{1/2}=d$\,.
Already the
inclusion of Fraunhofer scattering alone leads to a drastic reduction of the
Ehrenfest time: Injected electrons spread much faster over the whole cavity
than according to the estimate in Eq.~(\ref{eq:ehrtime}). An important
consequence of this modification is that the parameter 
$\widetilde{\tau}_E$ is now of
comparable magnitude or even smaller than the time $\tau_{0}$ 
beyond which universal behavior for $P(t)$ sets in 
[$P(t)\propto \exp{(-t/\tau_D)}$
for chaotic and $P(t)\propto (t/\tau_D)^{-\beta}$ for regular dots].  
In this regime system-specific deviations of the dwell time distribution from
a universal decay law are more pronounced 
than characteristic mean differences
between regular and chaotic cavities \cite{wirtz1,wirtz2,rotter1}.
For our estimate of the Fano factor $F$ we therefore take into
account
the exact dwell time distribution $P(t)$, resulting from the
quasi-classical Monte-Carlo trajectory simulation for the particular system
we study. Following the analysis \cite{sylvestrov}
based on the contribution of noiseless transmission 
channels, we find 
\begin{equation}\label{eq:integralfano}
F=1/4\,\left[1-\int_0^{\widetilde{\tau}_E}P(t)\,dt\right]=
1/4\,\int_{\widetilde{\tau}_E}^\infty P(t)\,dt\,.
\end{equation}
Note that Eq.~(\ref{eq:integralfano}) is applicable to 
chaotic and regular systems and is expected to be valid
irrespective of whether the origin of spreading is ballistic
scattering at the boundary or diffractive scattering inside of the
cavity.\\\begin{figure}[!t]
      \includegraphics[angle=0,width=85mm]{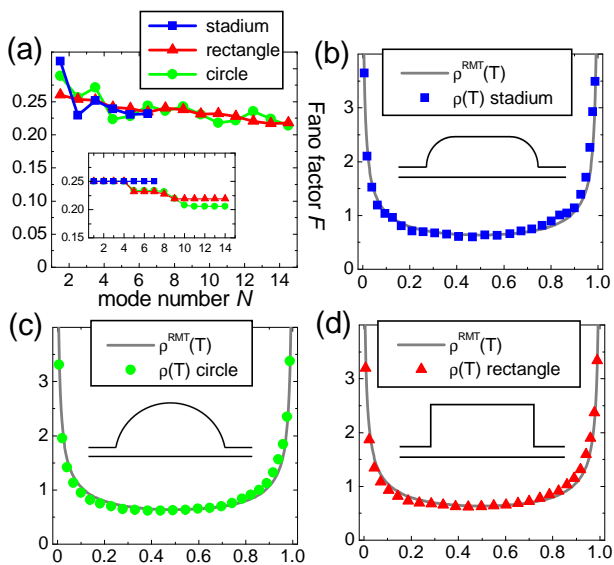}
      \caption{(a) Fano factor $F$ for scattering systems with
       chaotic (stadium) and regular (circle, rectangle) classical 
       dynamics as a function of $k_F$, in units of the
       number of open modes $N={\rm int} (k_Fb/\pi)$
       (each data point is averaged over 200 equidistant 
       $k_F$-values). 
       The inset shows the fit parameter free prediction 
       with diffractive corrections
       [see Eq.~(\ref{eq:integralfano})].
       (b)-(d) Distribution of transmission eigenvalues $\rho(T)$ averaged
       over the $k_F$-range depicted in (a). Remarkably the obtained values
       for $\rho(T)$ correspond very closely to the RMT-prediction
       $\rho^{\rm RMT}(T)$, irrespective of the chaoticity in the 
       cavity. The insets show the cavity geometries.}\label{fig:3}
\end{figure}It 
is now tempting to probe the convergence 
towards the RMT limit for $\widetilde{\tau}_E/\tau_D < 1$ also for the
distribution of transmission eigenvalues $\rho(T)$ and for other
regular and chaotic structures that feature only ballistic motion. 
We therefore investigate transport across the semi-circular, rectangular
and semi-stadium shaped cavity which are prototypical for regular and
chaotic ballistic motion, respectively (see insets of Fig.~\ref{fig:3}b,c,d).
Unlike in the case of the geometry of Fig.~\ref{fig:1}, 
where the number of open modes was fixed,
transport coefficients were calculated here for a wide range of $k_F$.
The dynamics of these cavities is classically scaling invariant. An increase
of $k_F$ thus probes directly the dependence of $\widetilde{\tau}_E/\tau_D$ on
$\hbar_{\rm eff}$, while keeping the classical phase space and $\Lambda$
fixed. Unlike the
  $\tau_D\rightarrow 0$ crossover performed with the help of the tunable
shutters as in the experiment \cite{oberholzer1} and in the simulation above, 
  the $k_F\rightarrow\infty$ limit would correspond to the ``pure''
  quantum-to-classical crossover as $\hbar_{\rm eff}\rightarrow
  0$.\\Employing the same high density of grid points we reach up to 15 
flux-carrying modes in the
semi-circular and rectangular billiard and 7 open lead modes in the
case of the semi-stadium \cite{rotter1}. We therefore can observe only the
onset
of the quantum-to-classical crossover for these three cavities. Their
area  $A=(4+\pi)/2$ and lead width $b=0.125$ are identical. 
Given this small leadwidth, i.e.~$b/\sqrt{A}=0.066$, 
our results agree very well with those shown in 
Fig.~\ref{fig:2} for small shutter openings: 
at all energies the Fano factor is close to $1/4$,
irrespective of the chaoticity in the cavity. 
However, for a higher number of open modes 
(i.e.~higher $k_F$) diffraction becomes less important since 
$\widetilde{\tau}_E $
becomes larger and a decrease in the Fano factor ensues, signifying the onset
of the quantum-to-classical crossover.
Our numerical data (Fig.~\ref{fig:3}a) agree indeed with this 
prediction. Moreover, the quantitative 
behavior is well described by Eq.\ (\ref{eq:integralfano}) 
(see inset of Fig.~\ref{fig:3}a).\\The RMT prediction for the distribution of
transmission eigenvalues 
 $T_n$ is given by \cite{jalabert}
$\rho^{\rm RMT}(T)=1/[\pi\sqrt{T(1-T)}]$.
Taking our original scattering geometry (Fig.~\ref{fig:1}) with a
setting of wide shutter openings (short $\tau_D$ or ``classical'') and no
disorder potential
(``regular'') we find a curve which clearly deviates from this 
RMT-prediction (see Fig.~\ref{fig:2}c,d).
Remarkably, the RMT-limit is, however, restored equally well by
either (1) further closing the shutters (longer $\tau_D$ or ``quantum'') or 
(2) by increasing the disorder potential (increase of $\langle\Lambda\rangle$
or ``chaotic''). In line with this observation  
the stadium-shaped, circular and rectangular cavity all fulfill the
$\rho^{\rm RMT}(T)$ remarkably well (see Fig.~\ref{fig:3}b,c,d) 
due to their small lead openings (i.e.~long dwell times). 
This suggests that for $\widetilde{\tau}_E /\tau_D\ll 1$ 
not just $F$ but also the distribution 
function $\rho(T)$ approaches the 
``universal'' RMT limit.\\To 
summarize, we have numerically verified the behavior of the Fano factor 
$F$ in a realistic scattering system with a tunable disorder 
potential and tunable shutters. We find that diffractive 
quantum scattering is sufficient 
to establish the RMT eigenvalue distribution $\rho^{\rm RMT}(T)$,
irrespective of regular or chaotic dynamics. The chaotic-to-regular and
quantum-to-classical crossover in $F$ can be estimated by a 
generalization of a previously proposed dependence \cite{sylvestrov}
on the Ehrenfest time $\widetilde{\tau}_E$ [Eq.\ (\ref{eq:integralfano})],
provided that the definition of the Ehrenfest time is properly 
modified to include diffraction.\\We
thank C.W.J.~Beenakker, J.~Cserti, V.A.~Gopar, F.~Libisch,
I.~Rotter, M.~Seliger, H.~Schomerus, E.V.~Sukhorukov and L.~Wirtz
for helpful discussions. Support by the Austrian Science
Foundation (Grant No.~FWF-SFB016 and No.~FWF-P17359) 
is gratefully acknowledged.

\end{document}